\documentclass[12pt]{article}
\usepackage{a4wide,graphicx}
\usepackage{cite}
\title{Searching for  exotic states in the $N \pi K $ system}
\author{
K. P. Khemchandani$^1$\footnote{kanchan@teor.fis.uc.pt}, A. Mart\'inez Torres$^2$\footnote{amartine@ific.uv.es}, and E.~Oset$^2$\footnote{oset@ific.uv.es} \\
{\small{\it $^1$ Centro de F\'isica Te\'orica, Departamento de F\'isica,}}\\
{\small{\it Universidade de Coimbra, P-3004-516 Coimbra, Portugal. 
}}\\
{\small{\it $^2$Departamento de F\'{\i}sica Te\'orica and IFIC,
Centro Mixto Universidad de Valencia-CSIC,}}\\
{\small{\it Institutos de
Investigaci\'on de Paterna, Aptdo. 22085, 46071 Valencia, Spain}}\\
}
\date{\today}
\begin{document}
\maketitle
\begin{abstract}
We  study the $N \pi K$ system in order to investigate the possibility of existence of strangeness +1 resonance(s). The formalism consists of solving  the Faddeev equations with the $N \pi$, $\pi K $ and $K N$ $t$-matrices obtained from chiral dynamics. 
The same formalism, which leads to the finding of several $1/2^+$ resonances in the corresponding three-body $S$ = -1 channels in the 1400-2000 MeV  energy region, results into only one broad bump around 1700 MeV with isospin  0. The amplitudes in isospin  1 and 2 configuration do not have any resonant structure. 
 \end{abstract}

\section{Introduction}

  The observation of a peak in the $K^+ n$ invariant mass for the 
  $\gamma n \to K^+ K^- n$ reaction on a
$^{12}C$ target at Spring8/Osaka \cite{penta} raised great hopes that for the
first time a strangeness S=1 narrow exotic baryon could be found. The peak was thus
associated to a pentaquark, since the standard $3q$ states cannot produce S=1.   
Subsequently, many experiments were done, some which reproduced this peak and others which did
not, and the issue stimulated a large number of theoretical works that gave a
huge impetus to the field of hadron structure (see the extensive list of references, for example, in  \cite{Nam:2004fh,Karliner:2004gr}). Waters calmed down, a thorough
experimental review  was written in \cite{Hicks:2005gp} and a period of rest
followed till a new experimental analysis was done at LEPS confirming the
original peak, now on a deuteron target and with more statistics 
\cite{Nakano:2008ee}. Although one cannot rule out an interpretation of the peak
as a consequence of the particular set up of LEPS, no alternative conventional 
explanation for this peak has been provided.

   On the theoretical side most of the works concentrated on finding possible states of
five quarks (pentaquark). From the perspective of the meson-baryon interaction the situation does not look encouraging, since the $KN$ interaction obtained from chiral Lagrangians is basically repulsive
in nature \cite{Bernard:1995dp}, and one does not expect to find a narrow (long lived) resonance,  as the one claimed in \cite{penta}, in this system. This is why very early there were suggestions that if the
peak represented a new state it could be a bound state of three hadrons, $K
\pi N$, with the pion acting as a glue between the nucleon and the $K$, which would only be bound by about 30 MeV. However,
investigations along this line,  weakly concluded the difficulty to have this
system as a bound state \cite{Bicudo:2003rw,LlanesEstrada:2003us}.

 The purpose of this paper is to perform a thorough calculation of the 
 $K \pi N$ system using Faddeev equations to see the possibility to find bound
 states or resonances. The interest in the three hadron systems is old. In 
 \cite{Harari:1964zz} there was already a study of such possible systems based
 only on symmetries. However, recently  a qualitative step forward has been taken in
 this topic, which has been made possible by combining elements of unitarized
 chiral perturbation theory $U\chi P T$
 \cite{Kaiser:1995eg,Dobado:1996ps,Oller:1997ti,Oller:1998hw, ollerulf,granada,osaka}
 with Faddeev equations in coupled channels \cite{MartinezTorres:2007sr}.
In \cite{MartinezTorres:2007sr,MartinezTorres:2008is}
  systems of two mesons and one baryon with strangeness $S=-1$ 
 were studied, finding resonant states
which could be identified with two $\Lambda$ and four $\Sigma$ known low-lying resonances with $J^P=1/2^+$. 
 Similarly, in the case of
the $S=0$ sector the $N^*(1710)$ appears neatly as a resonance of the $\pi \pi N$
system, as well as including the channels coupled to $\pi \pi N$ within SU(3)
\cite{Khemchandani:2008rk}. The study in $S$ = 0 sector
was further extended by using the experimental data on the $\pi N$ scattering and by adding more coupled channels in \cite{MartinezTorres:2008kh}, the outcome of which was the dynamical generation of three resonances, one with quantum numbers of  the $N^*$ (2100), another with those of the $\Delta$ (1910), plus a new $N^*$ at $\sim$ 1920 MeV (also predicted by Jido et. al. \cite{Jido}).

Developments along the same direction in three-meson system produced a
resonant state of $\phi K \bar{K}$ \cite{MartinezTorres:2008gy} which could be 
identified with the X(2175) resonance reported at BABAR 
\cite{Aubert:2006bu,Aubert:2007ur} and later on at BES \cite{:2007yt}.

The achievements obtained in the former studies and especially the finding of several low-lying $S$ =-1 
resonances with two meson-one baryon structure  motivates us to have a fresh look at the $\pi K N$ system using our formalism and
to  make a thorough
 investigation of the possibility to have the system bound. This is the
 purpose of the present paper. As we will show in the following section, we do not get the system bound in the region of the 
 possible S=1 state of  \cite{penta}. At higher energies a bump
 appears which, however, does not have the ordinary shape of the resonances that
 we have found in other channels. This could correspond to some of the bumps 
 seen using the time delay method in the analysis of the $K N $ system in 
\cite{Kelkar:2003pt}.

\section{Formalism and Results}
We solve the Faddeev equations for the $N \pi K$ system following the formalism developed in \cite{MartinezTorres:2007sr,MartinezTorres:2008is,Khemchandani:2008rk}, in which it was shown that the integral Faddeev equations can be rewritten as a coupled system of algebraic equations using unitary chiral dynamics.  The main feature of this approach  is the explicit cancellation between the contribution of the off-shell part of the two-body $t$-matrices in the three-body diagrams and the corresponding three-body contact term  originating from  the same chiral Lagrangians. Due to this peculiarity, the Faddeev equations can be transformed into the following set of coupled channel equations \cite{MartinezTorres:2007sr,Khemchandani:2008rk}
\begin{eqnarray} \nonumber
T^{\,12}_R&=&t^1g^{12}t^2+t^1\Big[G^{\,121\,}T^{\,21}_R+G^{\,123\,}T^{\,23}_R\Big] \\ \nonumber
T^{\,13}_R&=&t^1g^{13}t^3+t^1\Big[G^{\,131\,}T^{\,31}_R+G^{\,132\,}T^{\,32}_R\Big] \\ \nonumber
T^{\,21}_R&=&t^2g^{21}t^1+t^2\Big[G^{\,212\,}T^{\,12}_R+G^{\,213\,}T^{\,13}_R\Big] \\ \label{Trest}
T^{\,23}_R&=&t^2g^{23}t^3+t^2\Big[G^{\,231\,}T^{\,31}_R+G^{\,232\,}T^{\,32}_R\Big] \\ \nonumber
T^{\,31}_R&=&t^3g^{31}t^1+t^3\Big[G^{\,312\,}T^{\,12}_R+G^{\,313\,}T^{\,13}_R\Big] \\  \nonumber
T^{\,32}_R&=&t^3g^{32}t^2+t^3\Big[G^{\,321\,}T^{\,21}_R+G^{\,323\,}T^{\,23}_R\Big]  \nonumber
\end{eqnarray}
where $t^i$, $i=1,2,3$, represents the two-body $t$-matrices, which have been obtained by solving the Bethe-Salpeter equation in a coupled channel formalism. The required kernels, i.e., the  potentials which describe the interaction of the different pairs of the system are calculated using chiral Lagrangians \cite{Inoue:2001ip,Oset:1997it,Oller:1998hw}. In Eq. (\ref{Trest}), $g^{ij}$ is the  three-body Green's function of the system and 
$G^{ijk}$ is a loop function of three-particles (see \cite{MartinezTorres:2007sr,Khemchandani:2008rk} for more details). The matrices $t^i$, $g^{ij}$ and $G^{ijk}$ are all projected in $S$-wave, thus giving total $J^P=1/2^+$.

The $T^{ij}_{R}$ partitions consider all the different contributions to the three-body $T$- matrix in which the last interactions are given in terms of the two-body $t$-matrices $t^j$ and $t^i$, respectively. The $T^{ij}_{R}$ matrices are related to the Faddeev partitions $T^i$ through
\begin{equation}
T^i =t^i\delta^3(\vec{k}^{\,\prime}_i-\vec{k}_i) + \sum_{j\neq i=1}^3T_R^{ij}, \quad i=1,2,3
\end{equation}
where $\vec{k}_{i}$ ($\vec{k}^\prime_{i}$) is the initial (final) momentum of the particle $i$. Thus, the full three-body $T$-matrix is given by
\begin{eqnarray}
T&=&\sum_{i=1}^{3}T^{i}=\sum_{i=1}^{3}t^i\delta^3(\vec{k}^{\,\prime}_i-\vec{k}_i) +T_{R}\nonumber\\
T_{R}&\equiv&\sum_{i=1}^3\sum_{j\neq i=1}^{3}T^{ij}_{R}
\end{eqnarray}

 As our objective is to search for peaks in the $T$-matrix which can be associated with physical states, we can restrict ourselves to the study of the properties of \begin{equation}
T^*_R\equiv T_{R}- \sum_{i=1}^3\sum_{j\neq i=1}^{3}t^{i}g^{ij}t^{j}
\end{equation}
since neither $t^i\delta^3(\vec{k}^{\,\prime}_i-\vec{k}_i)$ nor the $t^{i}g^{ij}t^{j}$ terms can give rise to any three-body resonance. 

We study the $N \pi K $ system for total charge +1 taking into account four channels to solve Eqs. (\ref{Trest}): $p \pi^0K^0$, $n\pi^0K^+$, $p\pi^-K^+$, $n\pi^+ K^0 $.  In this case, the coupled channels appearing for the calculation of the two-body $t$-matrices are listed below:
\begin{itemize}
\item{$K^+\Sigma^0$, $K^0\Sigma^+$, $K^+\Lambda$, $\pi^0 p$,  $\pi^+ n$,  $\eta p $ for the $\pi N$ interaction with charge +1.}
\item{$K^+\Sigma^-$, $K^0\Sigma^0$, $K^0\Lambda$, $\pi^- p$, $\pi^0 n$, $\eta n $ for the $\pi N$ interaction with null charge.}
\item{$\pi^+K^0$, $\pi^0 K^+$ for $\pi K$ interaction with charge +1.}
\item{$\pi^-K^+$, $\pi^0 K^0$ for $\pi K$ interaction with charge 0.}
\item{And $K^0p$, $K^+n$ for $KN$ interaction with charge +1, $K^0n$ for charge 0 and $K^+p$  for charge +2.}
\end{itemize}
The meson-baryon potential obtained with chiral Lagrangians has the general form, after projecting in $S$-wave
\begin{equation}
V^{MB}_{ij} =-\frac{1}{4f^2}C_{ij}(2\sqrt{E}-M_{i}-M_{j})\sqrt{\frac{M_{i}+E_{i}(E)}{2M_{i}}}
\sqrt{\frac{M_{j}+E_{j}(E)}{2M_{j}}}
\end{equation}
where $f$ is the pion decay constant, $C_{ij}$ are coefficients which depend on the interaction under consideration, $E_{i}$ and $M_{i}$ ($E_{j}$ and $M_{j}$) are the energy and mass, respectively, of the incoming (outgoing) baryon and $E$ the total energy of the interacting particles. For the $\pi N
$ system and its coupled channels for total charge zero,  which dynamically generate the $N^*$ (1535), the coefficients $C_{ij}$ can be found in \cite{Inoue:2001ip}, while for the $K N$ system in \cite{Oset:1997it}.  The coefficients for the $\pi N$ system for total charge +1 are given in Table \ref{Ta} .

\begin{table}[h!]
\caption{$C_{ij}$ coefficients for the $\pi N$ interaction with charge +1}
\begin{center}
\begin{tabular}{c|cccccc}
\hline\hline
&$K^+\Sigma^0$&$K^0\Sigma^+$&$K^+\Lambda$&$\pi^0p$&$\pi^+n$&$\eta p$\\
\hline\\

$K^+\Sigma^0$&0&$\sqrt{2}$&0&-$\frac{1}{2}$&$\frac{1}{\sqrt{2}}$&-$\frac{\sqrt{3}}{2}$\\
\\
$K^0\Sigma^+$&&1&0&$\frac{1}{\sqrt{2}}$&0&-$\sqrt{\frac{3}{2}}$\\
\\
$K^+\Lambda$&&&0&-$\frac{\sqrt{3}}{2}$&-$\sqrt{\frac{3}{2}}$&-$\frac{3}{2}$\\
\\
$\pi^0p$&&&&0&$\sqrt{2}$&0\\
\\
$\pi^+n$&&&&&1&0\label{Ta}\\
\\
$\eta p$&&&&&&0\\
\hline\hline
\end{tabular}
\end{center}
\end{table}

The potential for the $\pi K$ system can be obtained from \cite{Oller:1998hw}, in which the $\kappa$(800) gets dynamically generated.

The $T^*_{R}$ matrix for different possible total isospins has  been obtained using the following relations:
\begin{eqnarray}
|N\pi K;I=0, I_{\pi K}=1/2\rangle&=&\frac{1}{\sqrt{6}}\Big[|p\pi^0K^0\rangle-\sqrt{2}|p\pi^-K^+\rangle+\sqrt{2}|n\pi^+K^0\rangle+|n\pi^0K^+\rangle\Big]\nonumber\\
|N\pi K;I=1, I_{\pi K}=1/2\rangle&=&\frac{1}{\sqrt{6}}\Big[|p\pi^0K^0\rangle-\sqrt{2}|p\pi^-K^+\rangle-\sqrt{2}|n\pi^+K^0\rangle-|n\pi^0K^+\rangle\Big] \\
|N\pi K;I=1, I_{\pi K}=3/2\rangle&=&\frac{1}{\sqrt{6}}\Big[\sqrt{2}|p\pi^0K^0\rangle+|p\pi^-K^+\rangle+|n\pi^+K^0\rangle-\sqrt{2}|n\pi^0K^+\rangle\Big]\nonumber\\
|N\pi K;I=2, I_{\pi K}=3/2\rangle&=&\frac{1}{\sqrt{6}}\Big[\sqrt{2}|p\pi^0K^0\rangle+|p\pi^-K^+\rangle-|n\pi^+K^0\rangle+\sqrt{2}|n\pi^0K^+\rangle\Big]\nonumber
\end{eqnarray}
and the phase convention $|\pi^+\rangle=-|I_{\pi}=1,I_{\pi z}=1\rangle$. The $I$ and $I_{\pi K}$ in the above equations represent the total isospin of the three-body system and that of the $\pi K$ system, respectively.

We calculate Eqs.(\ref{Trest}) as a function of the total energy and the invariant mass of the subsystem of particles $2$ and $3$. 
We denote these two variables as $\sqrt{s}$ and $\sqrt{s_{23}}$, respectively. The invariant masses of the other two subsystems can 
be written in terms of $\sqrt{s}$ and $\sqrt{s_{23}}$ as explained in \cite{MartinezTorres:2007sr}. We have calculated the 
$T_R$ matrices for the energy range 1500 MeV  $<\sqrt{s}<$  2100 MeV,  with the motivation to find some structure around 1540 MeV, and for 600 $<\sqrt{s_{23}}<$  1100 MeV to generate dynamically the $\kappa$ (800)  in the $KN$ subsystem in order to have some attractive interaction in the three-body system.

We do not find any resonance in the isospin 1 and 2 configurations. We obtain one peak with a broad structure in the squared amplitude in isospin zero (i.e., when the $\pi K$ subsystem is in isospin $1/2$) around 1720 MeV. The full width at half maximum of the peak is of the order of 200 MeV. These features have nothing in common with the resonance claimed in \cite{penta}. The value of $\sqrt{s_{23}}$,
for which the bump is found, is around the mass of the $\kappa$ (800) resonance. We show this peak in Fig.1, where we plot the amplitude square for the total isospin zero.  As is evident from the figure, the structure of the peak is far from a Breit-Wigner. Though the strength of this amplitude is similar to that of the corresponding $S=$ -1 case \cite{MartinezTorres:2007sr}, its shape is different from those of the clear resonances found in the latter.  Its unconventional peaking behaviour cast doubts whether this peak could have a pole associated in the complex plane, the accepted criterium to define a peak as a resonance. The technique to extrapolate the amplitudes to the complex plane with the two variables that we have is not available and looks nontrivial, since it involves working with complex momenta for some particles and real for others. Yet, independently from whether the structure found deserves or not to be called a resonance, the fact remains that the chiral dynamics of this coupled channel three-body system leads to such a bump in the cross section in a region where the system clusters like a $\kappa$ and a nucleon. This peak should be visible in $KN$ scattering with the quantum numbers $I=0$, $J^P=1/2^+$, but even better in the $KN\rightarrow \pi K N$ reaction, since the peak appears well above the $\pi K N$ threshold, or in any reaction producing $\pi KN$ in $I=0$ in the final state.
\begin{figure}[h!]
\begin{center}
\includegraphics[width=12cm]{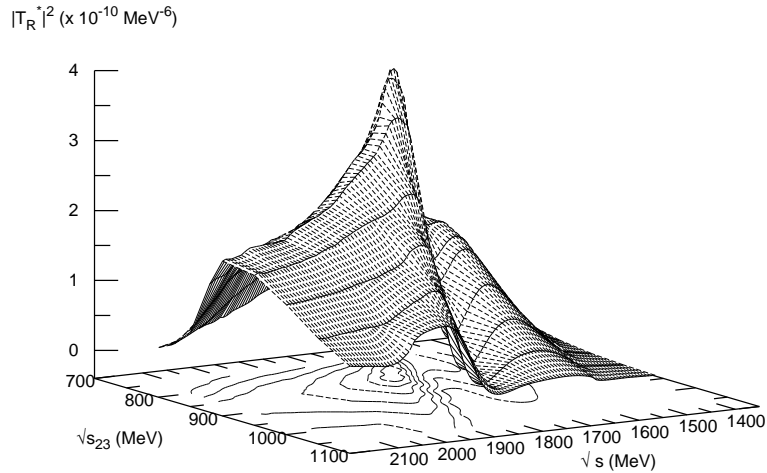}
\caption{The isospin zero amplitude squared for the $N\pi K$ system as a function of the total energy and the invariant mass of the $\pi K$ subsystem.}
\label{fig1}
\end{center}
\end{figure}

Interestingly, a broad structure at around 1800 MeV seems to be present in the data \cite{Hyslop:1992cs} of $K^+N$ scattering in the $P_{01}$ partial wave and in the time delay analysis of these data \cite{Kelkar:2003pt}, which could correspond to the peak shown in Fig.1. 
\section{Summary}
The possibility of existence of strangeness +1 baryon with a strong coupling to the $N \pi K $ system has been investigated by solving Faddeev equations in the formalism which has generated dynamically many strange and non strange resonances in three-body systems. 
We do not find any structure in the energy region close to 1542 MeV, therefore, the interpretation of a possible $\Theta^+$ as a  $N\pi K$ bound state, with all the interactions in $s$-wave, is ruled out. A bump is found around 1720 MeV with about 200 MeV of width and with isospin zero, which reveals the underlying chiral dynamics of the three-body system, and that we hope  can be seen in $K^+N$ scattering, but much better in reactions producing $\pi K N$ in $I=0$ in the final state. Our study should stimulate experimental work in this direction.

\section*{Acknowledgments}  
This work is partly supported by DGICYT contract number
FIS2006-03438. We acknowledge the support of the European Community-Research 
Infrastructure Integrating Activity
"Study of Strongly Interacting Matter" (acronym HadronPhysics2, Grant Agreement
n. 227431) under the Seventh Framework Programme of EU.  K. P. Khemchandani thanks the support by the Funda\c{c}\~{a}o para a Ci\^{e}ncia e a Tecnologia of the Minist\'erio da Ci\^{e}ncia, Tecnologia e 
Ensino Superior of Portugal (SFRH/BPD//2008). A. M. T is supported by a FPU fellowship of the Ministerio de Ciencia y Tecnolog\' ia.

\end{document}